\documentclass[fleqn,twoside]{article}

\usepackage{graphicx}
\usepackage{espcrc2}

\newcommand{\be}{\begin{equation}}
\newcommand{\ee}{\end{equation}}
\newcommand{\MeV}{\rm{MeV}}
\newcommand{\GeV}{\rm{GeV}}

\newcommand{\rhobar}{\overline {\rho}}
\newcommand{\etabar}{\overline{\eta}}
\newcommand{\Bo}{\rm{B^{0}}}
\newcommand{\Bob}{\overline{\rm{B}^{0}}}
\newcommand{\Ko}{\rm{K^{0}}}
\newcommand{\Kob}{\overline{\rm{K}^{0}}}
\newcommand{\dmd}{\Delta m_d}
\newcommand{\dms}{\Delta m_s}
\newcommand{\epsilonk}{\varepsilon_K}
\newcommand{\fbdsqbd}{f_{B_d} \hat B_{B_d}^{1/2}}
\newcommand{\fbssqbs}{f_{B_s} \hat B_{B_s}^{1/2}}
\newcommand{\sna}{\sin{2\alpha}}
\newcommand{\snb}{\sin{2\beta}}

\title{The role of Lattice QCD in flavor physics}

\author{V. Lubicz\address{Dipartimento di Fisica, Universit\`a di Roma Tre and
INFN, Sezione di Roma III \\ Via della Vasca Navale 84, I-00146 Rome, Italy}}
       
\begin{document}

\begin{abstract}
Understanding flavor physics is one of the most important tasks of particle 
physics today, which is motivating an extraordinary experimental and theoretical
investigational effort. Important progress in this field has already been 
achieved in the last few years, with Lattice QCD calculations playing an 
essential role in this effort. I will describe some lattice contributions to the
studies of flavor physics by focusing particularly on the determination of the 
CKM matrix and on the study of CP violation in the Standard Model.
\vskip -1em
\end{abstract}

\maketitle

\section{INTRODUCTION}
Flavor physics is the subject of intense expe\-ri\-mental and theoretical 
research. There are well founded reasons for that, which can be summarized as 
follows:

\vspace{0.2truecm}\noindent
-- Flavor physics is (well) {\em described} but not {\em explained} in the 
Standard Model (SM). This is ma\-ni\-fested by the large number of free 
parameters in the flavor sector of the SM (10 parameters in the quark sector 
only). In addition, many of these parameters, which in the SM Lagrangian enter 
as Yukawa couplings $y_{ij}$, assume unnaturally small values (at least in the 
mass diagonal basis). The fermion masses $m_i \sim y_{ii} v$, unlike the gauge 
boson masses $M_W$ and $M_Z$, are much smaller than the electroweak symmetry 
breaking scale $v$, with the notable exception of the top quark mass. 

Besides providing us with an accurate parametric description, the SM does not 
explain flavor physics: we do not know why fermions come in families, nor what 
generates the hierarchy of masses, nor the quark mixing pattern encoded in the
CKM matrix. In the leptonic sector, even the basic structure of the neutrino 
mass and mixing matrix is not established yet.

\vspace{0.2truecm}\noindent
-- In the absence of gauge symmetry breaking all particles in the SM remain 
massless and flavor symmetry is exact. The breaking of flavor symmetry (i.e. the
generation of fermion masses) is originated by gauge symmetry breaking. The two 
mechanisms seem to be closely related and it may not be a coincidence that a 
definite explanation for both of them is lacking at present.

\vspace{0.2truecm}\noindent
-- Understanding the origin of CP violation is a fundamental issue in particle 
physics. This phenomenon has played a significant role in the primordial 
evolution of the Universe and it is expected to explain the origin of the 
observed asymmetry between matter and antimatter in the Universe. In the SM, the
only source of CP violation is a single (physical) complex phase in the quark 
mass matrix. To verify the SM mechanism of CP violation is thus one of the most 
important tasks of current studies of flavor physics.

\vspace{0.2truecm}\noindent
-- It is beyond any reasonable doubt that the SM only represents the low energy 
limit of a more fundamental theory. Besides the well known conceptual problems 
of the SM, the most obvious of which is the non inclusion of a quantum 
descri\-ption of gravity, there are important phenomenological indications of
New Physics, i.e. physics beyond the SM, that have manifested in the last few 
years. The following is a (possibly incomplete) list of such indications. The 
existence of neutrino masses is experimentally well established, and the 
measured values of neutrino mass diffe\-ren\-ces point to an energy scale which 
is close to the Grand Unification scale. Unification of couplings itself is 
indicated but not precisely achieved in the SM, thus suggesting the existence of
New Physics at intermediate energy scales (below the GUT scale). At a 
cosmological level we find that va\-cu\-um energy represents a significant 
fraction of the total energy density of the Universe ($\Omega_{vac} \sim 0.7$) 
but this value is extraordinary small on a particle physics energy scale, a 
phenomenon for which we have no convincing explanation. The rest of the energy 
density is mostly provided by matter ($\Omega_{mat} \sim 0.3$), but the largest
fraction of it (more than 80\%) is not of baryonic nature. Candidates for this 
dark matter are absent in the SM. Finally, the SM does not provide an acceptable
explanation for baryogenesis (i.e. matter-antimatter asymmetry): the departure 
from thermal equili\-brium is not sufficiently strong in the phase transition 
generated by electroweak symmetry brea\-king. Even most notably, the amount of 
CP violation originated by the CKM mechanism in the SM is not enough to explain 
baryogenesis.

It is likely that further insights in New Physics will eventually come from 
studying flavor physics. A simple naturalness argument in the SM, based on the 
requirement that radiative corrections to the Higgs boson mass should not exceed
the upper bound on this mass deduced from electroweak precision tests, suggests
that New Physics should appear at an energy scale of ${\cal O}$(1~TeV). This 
scale is remarkably close to the region already explored by present experiments.
On the other hand, upper bounds on New Physics coming from the flavor 
sector, e.g. from the value of the neutral kaon mass difference $\Delta M_K$, 
point to a much higher e\-ner\-gy scale for New Physics, of ${\cal O}$(100~TeV).
The difficulty of explaining such a higher scale is sometimes called the 
{\it flavor problem}.

The above considerations illustrate the phenomenological and theoretical
importance of the study of flavor physics. On the experimental side, both 
high-energy collision experiments and high-statistics dedicated flavor factories
are providing us with results of increasing accuracy. In many relevant cases 
this accuracy has reached the level of 1\% or better. On the theory side, one
needs to match the precision achieved by experiments, which requires in 
particular an accurate determination of the theoretical input parameters. This 
is the present task of Lattice QCD calculations. In this talk I would like to 
show that we are already facing this task, and that Lattice QCD calculations
play an essential role in present studies of flavor physics. I will focus the
discussion on two main topics: i) the determination of the Cabibbo angle and ii)
the analysis of the Unitarity Triangle and CP violation. An important 
contribution of Lattice QCD to the studies of flavor physics is represented also
by the determination of quark masses, a topic which is reviewed by P.~Rakow at 
this conference~\cite{rakow}. Closely related to the subject of this discussion 
are the talk of M.~Wingate~\cite{wingate} on the status of Lattice Flavor 
Physics and the stimulating ``Experimenter's view of Lattice QCD" presented us 
by I.~Shipsey~\cite{shipsey}.

\section{FIRST ROW UNITARITY}
The determination of the Cabibbo angle is of particular phenomenological and
theoretical interest since it provides at present the most stringent unitarity 
test of the CKM matrix. This is expressed by the ``first row" unitarity 
condition:
\be
|V_{ud}|^2 + |V_{us}|^2 + |V_{ub}|^2 =1 \,.
\label{eq:1strow}
\ee
Since $|V_{ub}|\sim 10^{-3}$, its contribution to the Eq.~(\ref{eq:1strow}) can 
be safely neglected. 

The value of $|V_{ud}|$ is accurately determined from nuclear superallowed 
$0^+\to 0^+$ and neutron beta decays. An updated analysis of these results leads
to the very precise estimate~\cite{marciano} 
\be
\label{eq:vud}
|V_{ud}| = 0.9740 \pm 0.0005 \,,
\ee
where the average is dominated by superallowed transitions. In the following, I 
will therefore concentrate on the remaining entry $|V_{us}|$.

\subsection{Present status of $\mathbf{|V_{us}|}$}
The most accurate determination of $|V_{us}|$ is obtained from semileptonic kaon
decays ($K_{\ell3}$). The analysis of the experimental data gives access to the 
quantity $|V_{us}| \cdot f_+(0)$, where $f_+(0)$ is the vector form factor at 
zero four-momentum transfer square. In the SU(3) limit, vector current 
conservation implies $f_+(0)=1$. The deviation of $f_+(0)$ from unity represents
the main source of theoretical uncertainty. This deviation has been estimated 
many years ago by Leutwyler and Roos (LR)~\cite{LR}, who combined a leading 
order analysis in chiral perturbation theory (ChPT) with a quark model 
calculation. They obtained $f_+^{K^0 \pi^-}(0) = 0.961 \pm 0.008$, and this 
value still represents the referential estimate~\cite{PDG}.

By averaging old experimental results for $K_{\ell 3}$ decays with the recent 
measurement by E865 at BNL~\cite{E865}, and using the LR determination of the
vector form factor, the PDG quotes $|V_{us}| = 0.2200 \pm 0.0026$~\cite{PDG}. 
This value, once combined with the determination of $|V_{ud}|$ given in 
Eq.~(\ref{eq:vud}), implies about $2 \,\sigma$ deviation from the CKM unitarity 
condition, i.e. $|V_{us}|^{\rm unit.}\simeq \sqrt{1-|V_{ud}|^2} = 0.2265 \pm 
0.0022$. 

\begin{figure}[t]
\vspace{-0.5truecm}
\begin{center}
\includegraphics[width=7.5cm]{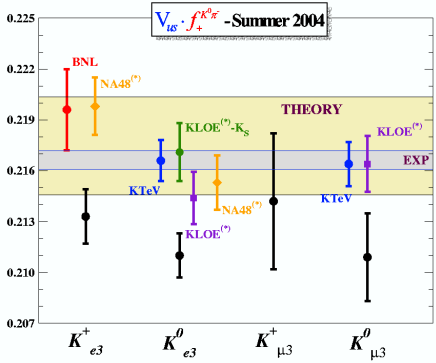}
\end{center}
\vspace{-1.5truecm}
\caption{\it Experimental results for $|V_{us}|\cdot f_+(0)$. The ``EXP" and 
``THEORY" bands indicate respectively the average of the new experimental 
results and the unitarity prediction combined with the LR and lattice (see 
Sect.\ref{sec:kl3}) determination of the vector form factor.}
\label{fig:fvus}
\vspace{-0.8truecm}
\end{figure}

With respect to the PDG analysis, however, a significant novelty is represented 
by several new experimental results, for both charged and neutral $K_{\ell 3}$ 
decays, which have been recently presented by KTeV \cite{KTEV}, NA48 \cite{NA48}
and KLOE \cite{KLOE}. Expressed in terms of $|V_{us}|\cdot f_+(0)$, these 
determinations are shown in Fig.~\ref{fig:fvus}, together with the BNL 
result and the averages of the old $K_{\ell 3}$ results quoted by the PDG.
Remarkably, the average of the new results~\cite{Mescia}, represented by the 
darker band in the plot (``EXP"), is in very good agreement with the unitarity 
prediction, once the LR determination of the vector form factor is taken into 
account. The unitarity prediction is shown in Fig.~\ref{fig:fvus} by the 
lighter band (``THEORY").

On the theoretical side, two important contributions to the determination of 
$|V_{us}|$ come from lattice calculations and have been presented at this 
conference~\cite{bernardproc,mesciaproc}. They concern leptonic and semileptonic
kaon decays respectively. That already shows that Lattice QCD can contribute to 
precision calculations in flavor physics. 

\subsection{$\mathbf{|V_{us}|}$ from leptonic kaon decays}
As a part of their extensive study of partially quenched QCD with three 
dynamical flavors of improved staggered quarks, the MILC Collaboration has 
presented the final results for the light pseudoscalar decay 
constants~\cite{bernardproc,milc},
\begin{eqnarray}
\label{eq:fmilc}
f_\pi & = & 129.5 \pm 0.9 \pm 3.6 ~\MeV \,, \nonumber \\
f_K   & = & 156.6 \pm 1.0 \pm 3.8 ~\MeV \,, \\
f_K/f_\pi & = & 1.210 \pm 0.004 \pm 0.013 \,, \nonumber
\end{eqnarray}
where the first error is statistical and the second systematic. The lattice data
have been fit to staggered chiral perturbation theory expressions and 
extrapolated in quark masses, lattice spacing and lattice volume. 

The uncertainties quoted in Eq.~(\ref{eq:fmilc}) represent an accuracy which has
no precedent in lattice calculations of these quantities. It should be mentioned
that there is a potential source of syste\-ma\-tic error which is not quoted in 
Eq.~(\ref{eq:fmilc}) coming from the so called ``fourth-root" trick applied to 
the staggered fermion determinant. A possibility exists that physical 
non-localities induced by this trick will remain in the continuum 
limit~\cite{milc}.

The precision achieved in Eq.~(\ref{eq:fmilc}) for $f_K/f_\pi$ allows for an 
accurate determination of $|V_{us}|/|V_{ud}|$ from the ratio of leptonic kaon
decay rates:
\be
\frac{\Gamma(K\to\mu \bar\nu_\mu(\gamma))}{\Gamma(\pi\to\mu\bar\nu_\mu (\gamma))
} = K \frac{|V_{us}|^2 f^2_K m_K \left(1-\frac{m^2_\mu}{m^2_K}\right)^2} 
{|V_{ud}|^2 f^2_\pi m_\pi \left(1-\frac{m^2_\mu}{m^2_\pi}\right)^2} \,,
\label{eqthirteen} 
\ee
where $K=0.9930(35)$ takes into account radiative corrections 
\cite{marciano-Vus}. By combining the experimental result $\Gamma(K\to\mu \bar
\nu_\mu(\gamma))/\Gamma(\pi\to\mu\bar\nu_\mu (\gamma)) = 1.334 (4)$ with the 
Lattice QCD determination of $f_K/f_\pi$ in Eq.~(\ref{eq:fmilc}) and the value 
of $|V_{ud}|$ given in Eq.~(\ref{eq:vud}), one obtains
\be
\label{eq:vuslept} 
|V_{us}| = 0.2219(1)(3)(4)(26) \,.
\ee
The errors come respectively from the uncertainties on $|V_{ud}|$, the 
experimental measurement of leptonic decay rates, the radiative corrections 
in Eq.~(\ref{eqthirteen}) and the lattice determination of $f_K/f_\pi$. The
estimate of $|V_{us}|$ in Eq.~(\ref{eq:vuslept}) is consistent with unitarity 
at the $1.4\,\sigma$ level. Unfortunately, it will be difficult to further 
reduce the lattice uncertainty which clearly dominates the error. 

\subsection{$\mathbf{|V_{us}|}$ from semileptonic kaon decays}
\label{sec:kl3}
The other theoretical progress in the determination of $|V_{us}|$ comes from the
studies of semileptonic $K_{\ell 3}$ decays and it is represented by the first 
(quenched) lattice determination with significant accuracy of the vector form 
factor at zero momentum transfer square $f_+(0)$~\cite{mesciaproc,our}. The 
lattice result turns out to be in very good agreement with the quark model 
estimate obtained by LR, thus putting the evaluation of this form factor on a 
firmer theoretical basis. Before outlining the strategy of the lattice 
calculation, whose details have been presented by F.~Mescia at this 
con\-fe\-rence~\cite{mesciaproc}, I would like to summarize the theore\-ti\-cal
status of the $f_+(0)$ evaluations.

A good theoretical control on $K_{\ell 3}$ transitions is assured by the 
Ademollo-Gatto (AG) theorem~\cite{ag}, which states that $f_+(0)$ is 
renorma\-li\-zed only by terms of at least second order in the breaking of 
SU(3)-flavor symmetry. Nevertheless, the error on the shift of $f_+(0)$ from 
unity represents not only the main source of theoretical uncertainty but it 
also dominates the overall error in the determination of $|V_{us}|$.

The amount of SU(3) breaking due to light quark masses can be investigated 
within ChPT, by performing an expansion of the form $f_+(0) = 1 + f_2 + f_4 + 
\ldots$, where $f_n = {\cal{O}}(p^n)={\cal{O}}[{M^n_{K,\pi}} / (4\pi f_\pi)^n]$.
Thanks to the AG theorem, the first non-trivial term in the chiral expansion, 
$f_2$, does not receive contributions of local operators appearing in the 
effective theory and can be computed unambiguously in terms of $M_K$, $M_\pi$ 
and $f_\pi$ ($f_2 = -0.023$, in the $K^0 \to \pi^-$ case \cite{LR}). The 
higher-order terms of the chiral expansion involve instead the coefficients of
local chiral operators, that are difficult to estimate. The quark model 
calculation by LR provides an estimate of the next-to-leading correction $f_4$, 
and it is based on a general parameterization of the SU(3) breaking structure 
of the pseudoscalar meson wave functions. 

An important progress in this study is re\-pre\-sented by the complete 
two-loop ChPT calculation of $f_4$, performed in Refs.~\cite{post,BT}. The 
result can be written in the form
\be
f_4 = \Delta(\mu) - \frac{8}{F_\pi^4} \left[ C_{12}(\mu) + C_{34}(\mu) 
\right] \left(M_K^2-M_\pi^2\right)^2 ,
\label{eq:f4}
\ee
where $\Delta(\mu)$ represents the loop contribution, expressed in terms of 
chiral logs and the ${\cal O}(p^4)$ low-energy constants, while the second term 
is the analytic one. As can be seen from Eq.~(\ref{eq:f4}), the local
contribution involves a single combination of two (unknown) chiral coefficients 
entering the effective Lagrangian at ${\cal O}(p^6)$. In addition, the 
separation between non-local and local contribution quantitatively depends 
on the choice of the renormalization scale $\mu$, only the whole result for 
$f_4$ being scale independent. This dependence is found to be large~\cite{cnp}; 
for instance, at three typi\-cal values of the scale one finds
\be
\label{eq:deltamu}
\Delta(\mu) = \left\{ 
\begin{array}{ccc}
0.031\,, & 0.015\,, & 0.004 \\
\mu = M_\eta & \mu = M_\rho & \mu = 1 \,\GeV
\end{array}
\right. \,.
\ee
An important observation by Bijnens and Tala\-ve\-ra \cite{BT} is that the 
combination of low-energy constants entering $f_4$ could be in principle 
constrained by experimental data on the slope and curvature of the scalar form 
factor. The required level of experimental precision, however, is far from what
is currently achieved. Thus, one is left with either the LR result or other 
model dependent estimates of the local term in Eq.~(\ref{eq:f4}). Recent 
attempts in this direction include the estimate by resonance saturation obtained
in Ref.~\cite{cnp} and the dispersive analysis of Ref.~\cite{jamin}. On the 
other hand, the large scale dependence of the ${\cal{O}}(p^6)$ loop calculation 
shown in Eq.~(\ref{eq:deltamu}) seems to indicate that the error $\pm 0.010$ 
quoted in Refs.~\cite{BT}-\cite{jamin} might be underestimated. A first 
principle lattice determination of the vector form factor is thus of great 
phenomenological relevance.
\begin{figure*}[t]
\vspace{-0.5truecm}
\begin{center}
\begin{tabular}{ccc}
\includegraphics[width=4.5cm]{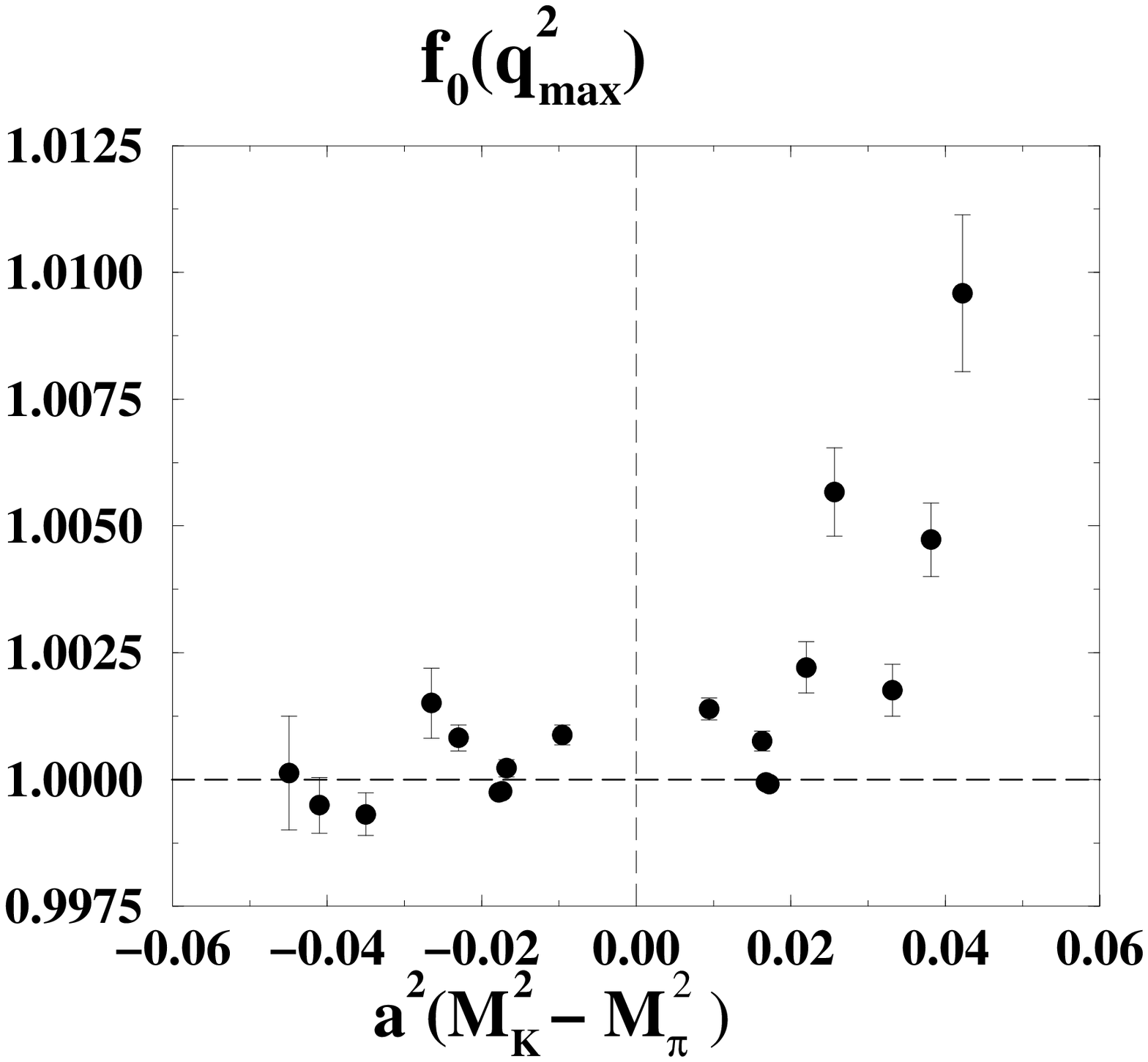} &
\includegraphics[width=4.65cm]{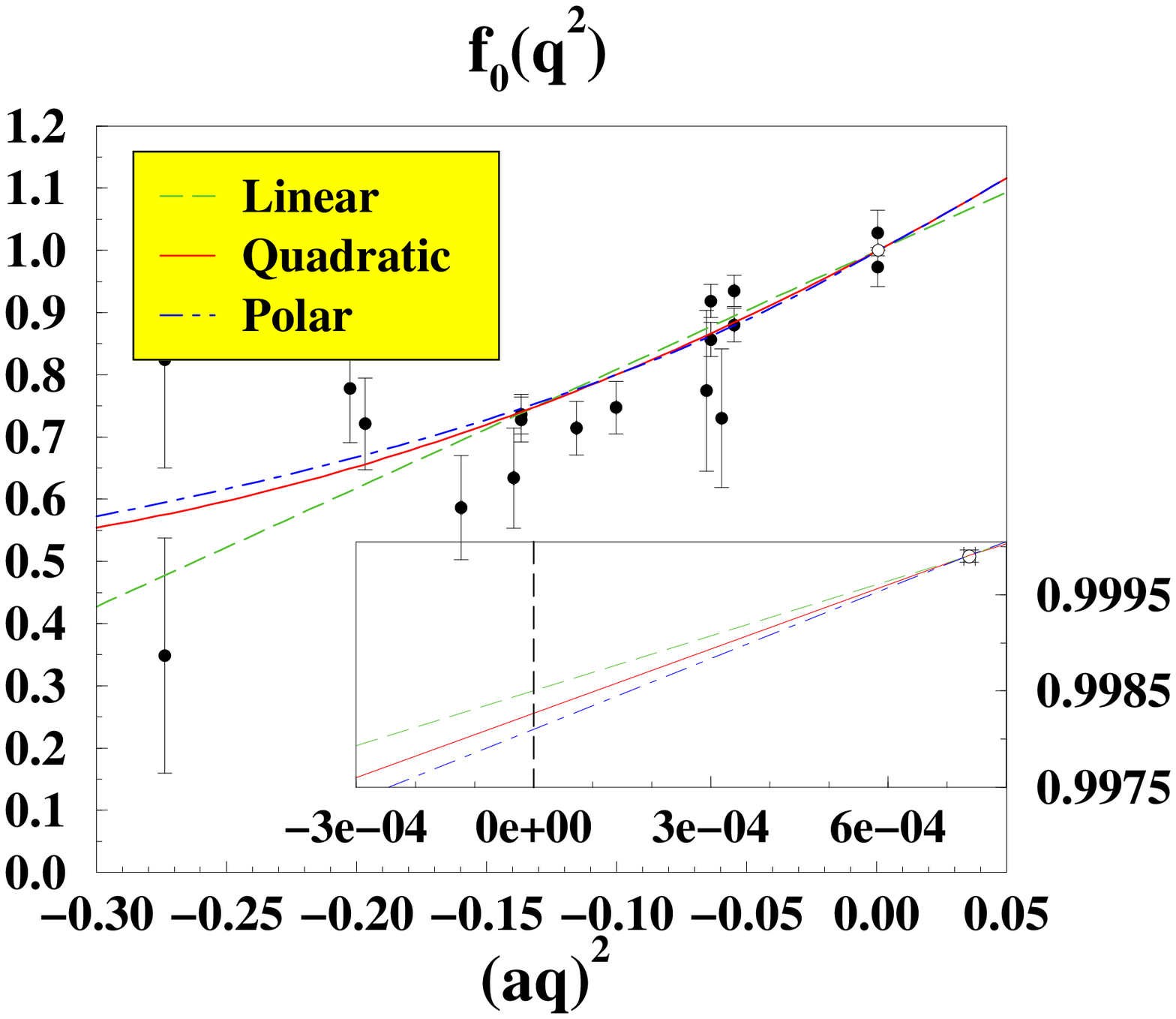} &
\includegraphics[width=4.2cm]{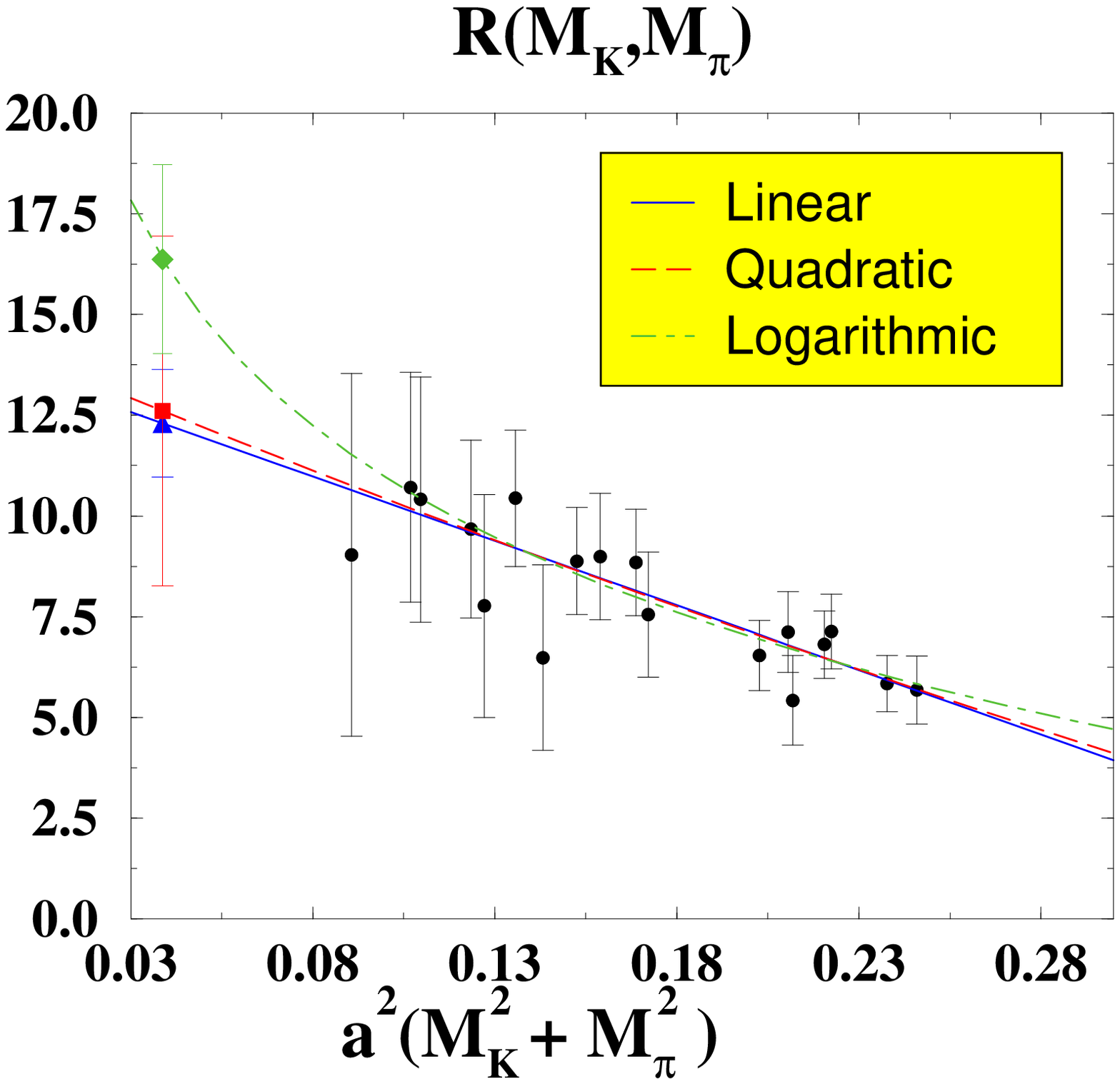}
\end{tabular}
\end{center}
\vspace{-1.2truecm}
\caption{\it {\rm Left:} Values of $f_0(q_{\protect\rm max}^2)$ versus the 
SU(3)-breaking parameter $a^2 (M_K^2 - M_{\pi}^2)$. {\rm Center:} The form 
factor $f_0(q^2)$ as a function of $q^2$ for one of the quark mass combinations 
used in the simulation. The inset is an enlargement of the region around 
$q^2=0$. {\rm Right:} Comparison among linear, quadratic and logarithmic 
chiral extrapolations of the ratio $R$ of Eq.~(\ref{eq:erre}) as a function
of $a^2 (M_K^2 + M_{\pi}^2)$.}
\label{fig:kl3}
\vspace{-0.2truecm}
\end{figure*}

In order to reach the challenging goal of about 1\% error on the lattice
determination of $f_+(0)$, a new strategy has been proposed and applied in the 
quenched approximation~\cite{our}. This strategy is based on three steps.

\vspace{0.2truecm}\noindent
{\bf 1) Precise evaluation of $\mathbf{f_0(q^2)}$ at $\mathbf{q^2= q_{\rm max}
^2}$}\par\noindent
This evaluation follows a procedure originally proposed by the FNAL group to 
study heavy-light form factors~\cite{FNAL}. For $K_{\ell 3}$ decays, the scalar 
form factor $f_0(q^2)$ can be calculated very efficiently at $q^2 = q_{\rm max}
^2 = (M_K - M_\pi)^2$ by studying the following double ratio of matrix elements,
\be 
\frac{\langle \pi 
| \bar{s} \gamma_0 u | K \rangle \, \langle K | \bar{u} \gamma_0 s | 
\pi \rangle}{\langle K | \bar{s} \gamma_0 s | K \rangle \, \langle \pi | 
\bar{u} \gamma_0 u | \pi \rangle} = C_{K,\pi}\,[f_0(q_{\rm max}^2)]^2 \,,
\label{eq:fnal2}
\ee
where all the external particles are taken at rest and $C_{K,\pi}=(M_K+M_\pi)^2
/(4 M_K M_\pi)$. There are several crucial advantages in using the double ratio 
(\ref{eq:fnal2}) which are described in details in Ref. \cite{our}. The most
important point is that this ratio gives values of $f_0(q_{\rm max}^2)$ with a 
statistical uncertainty smaller than 0.1\%, as shown in Fig.~\ref{fig:kl3} 
(left).

\vspace{0.2truecm}\noindent
{\bf 2) Extrapolation of $\mathbf{f_0(q_{\rm max}^2)}$ to $\mathbf{f_0(0) = 
f_+(0)}$}\par\noindent
This extrapolation is performed by studying the $q^2$-dependence of $f_0(q^2)$.
New suitable double ratios are introduced also in this step, that improve the 
statistical accuracy of $f_0(q^2)$. The qua\-li\-ty of the extrapolation is 
shown in Fig.~\ref{fig:kl3} (center). Three different functional forms in $q^2$ 
have been considered, namely a polar, a linear and a quadratic one. The 
differences among the results are considered in the evaluation of the systematic
error.

\vspace{0.2truecm}\noindent
{\bf 3) Extrapolation to the physical masses}\par\noindent
The physical value of $f_+(0)$ is finally reached after extrapolating the 
lattice results to the physical kaon and pion masses. The problem of the chiral
extrapolation is substantially simplified if the AG theorem 
is taken into account and if the lea\-ding (quenched) chiral logs are 
subtracted. This is achieved by introducing the following ratio
\be
R = \frac{\Delta f}{(M_K^2-M_\pi^2)^2} = \frac{ 1 + f_2^q -f_+(0) }
{(M_K^2-M_\pi^2)^2} \,,
\label{eq:erre}
\ee
where $f_2^q$ represents the leading chiral contribution calculated in quenched 
ChPT~\cite{our} and the quadratic dependence on $(M_K^2-M_\pi^2)$, driven by the
AG theorem, is factorized out. It
should be emphasized that the subtraction of $f_2^q$ in Eq.~(\ref{eq:erre}) 
does not imply necessarily a good convergence of (quenched) ChPT at ${\cal O}
(p^4)$ for the meson masses used in the lattice simulation. The aim of the 
subtraction is to access directly on the lattice the quantity $\Delta f$, 
defined in such a way that its chiral expansion starts at ${\cal O}(p^6)$ 
independently of the values of the meson masses. After the subtraction of 
$f_2^q$, the ratio $R$ of Eq.~(\ref{eq:erre}) is smoothly extrapolated in the 
meson masses as illustrated in Fig.~\ref{fig:kl3} (right). In order to check 
the stability of the extrapolation, linear, quadratic and logarithmic fits 
have been considered. The leads to the final result
\be
f_+^{K^0\pi^-}(0) = 0.960 \pm 0.005_{\rm stat} \pm 0.007_{\rm syst}\,,
\label{eq:f0final}
\ee
where the systematic error does not include quenching effects beyond ${\cal{
O}}(p^4)$. Removing this error represents one of the major goal of future 
lattice studies of $K_{\ell3}$ decays. 

The value (\ref{eq:f0final}) compares well with the LR result $f_+^{K^0 \pi^-}
(0) = 0.961 \pm 0.008$ quoted by the PDG~\cite{PDG} and, once combined with the 
average of the more recent experimental results, implies 
\be
|V_{us}|=0.2256 \pm 0.0022 \,,
\ee
in good agreement with unitarity. 

A strategy similar to the one discussed above has been also applied to study 
hyperon semileptonic decays on the lattice, and preliminary results have been 
presented at this conference~\cite{iperoni}.

\section{LATTICE QCD AND THE UNI\-TA\-RI\-TY TRIANGLE ANALYSIS}
Now, I would like to illustrate the role of Lattice QCD calculations in 
providing the input parameters to the analysis of the Unitarity Triangle (UT) 
and of CP violation. This study provides one of the most significant tests of 
flavor physics in the SM and a unique opportunity for searching New Physics.

\subsection{Lattice input parameters}
The most precise determination of the UT parameters is obtained by using
semileptonic B-decays, $\Bo-\Bob$ oscillations and CP asymmetries in the kaon 
and in the B sectors. This will be referred to as the ``standard 
analysis"~\cite{ref:noi1}-\cite{ref:loro2}, and relies on the following five
measurements: the ratio $\left | V_{ub} \right |/\left | V_{cb} \right|$, the
mass differences in the neutral B-meson systems, $\Delta {m_d}$ and the limit on
$\Delta {m_s}$, and the CP-violating quantities in the kaon ($\epsilonk$) and in
the B ($\snb$) sectors. Lattice QCD calculations play a central role by 
determining many of the theoretical inputs. A list of the most re\-le\-vant 
input quantities, with corresponding central values and errors, is given in 
Table~\ref{tab:inputs}\footnote{See {\tt http://www.utfit.org} for the complete
collection of input quantities and results.}.
\begin{table}[t]
\caption {\it Central values and errors of some of the most re\-le\-vant input 
quantities used in the UT analysis.}
\label{tab:inputs}
\begin{center}
\begin{tabular}{ll}
\hline\hline
 Parameter &  Value  \\ \hline\hline
$\lambda$  &  $0.2265  \pm  0.0020$ \\ \hline
$\left |V_{cb} \right |$(excl.)  & $ (42.1 \pm 2.1) \times 10^{-3}$  \\
$\left |V_{cb} \right |$(incl.)  & $ (41.4 \pm 0.7 \pm 0.6)\times 10^{-3}$  \\ 
$\left |V_{ub} \right |$(excl.)  & $ (33.0 \pm 2.4 \pm 4.6)\times 10^{-4}$  \\  
$\left |V_{ub} \right |$(incl.-LEP) & $ (40.9 \pm 6.2 \pm 4.7)\times 10^{-4}$\\
$\left |V_{ub} \right |$(incl.-HFAG)& $ (45.7 \pm 6.1) \times 10^{-4}$ \\ \hline
$\Delta m_d$  & $(0.503 \pm 0.006)~\mbox{ps}^{-1}$  \\
$\Delta m_s$  & $>$ 14.5 ps$^{-1}$ at 95\% C.L. \\
              & sensitivity 18.3 ps$^{-1}$      \\ 
$\snb$        & $ 0.739 \pm 0.048 $  \\ \hline
$\fbssqbs$    & $ 276 \pm 38 $ MeV   \\
$\xi=\fbssqbs/\fbdsqbd$ & $1.24 \pm 0.04 \pm 0.06$  \\
$\hat B_K$    & $0.86 \pm 0.06 \pm 0.14$  \\
\hline\hline
\end{tabular} 
\end{center}
\vspace{-1.0cm}
\end{table}
The main input which come from lattice calculations are the form factors
controlling semileptonic B-decays, the hadronic parameters $\fbssqbs$ and $\xi$
entering the amplitudes of $\Bo-\Bob$ oscillations, and the bag parameter $\hat 
B_K$ which parameterizes $\Ko-\Kob$ mixing. The central values and errors quoted
in Table~\ref{tab:inputs} are those adopted at the end of the ``CKM Unitarity 
Triangle'' workshops \cite{ref:ckm1,ref:ckm2} and by the HFAG~\cite{ref:hfag}.
With respect to previous analyses~\cite{ref:noi1,ref:noi2}, the novelties in 
Table~\ref{tab:inputs} are the final LEP/SLD likelihood for B$_s$ oscillations, 
the value of $|V_{ub}|$ from inclusive semileptonic decays \cite{ref:hfag}, the 
new value of $\snb$ and a new treatment of the non-perturbative QCD parameters 
controlling $\Bo-\Bob$ mixing, as explained below.

In previous analyses, the constraints coming from the experimental informations
on $\Delta m_d$ and $\Delta m_s$ were implemented as
\[
\Delta m_d \,\propto\, |V_{td}|^2 ~ f_{B_d}^2 \hat B_{B_d} 
\,\propto \,[(1-\rhobar)^2+\etabar^2] ~ f_{B_d}^2 \hat B_{B_d} 
\]
\vspace{-0.3cm}
\be
\Delta m_s \, \propto \, |V_{ts}|^2 ~ f_{B_s}^2 \hat B_{B_s}  
\,\propto\, f_{B_d}^2 \hat B_{B_d} \times \xi^2 
\ee
In this case, the input quantities are $\fbdsqbd$ and $\xi$. The hadronic 
parameter that is better determined from lattice calculations, however, is 
$\fbssqbs$, whereas $\xi$ and $\fbdsqbd$ are affected by larger uncertainties 
coming from the chiral extrapolations. These uncertainties are strongly
correlated. For this reason, a better approach consists in writing the 
constraints as follows:
\begin{eqnarray}
\Delta m_d &\propto& [(1-\rhobar)^2+\etabar^2] ~ f_{B_s}^2 \hat B_{B_s}/\xi^2 
\nonumber \\
\Delta m_s &\propto&  f_{B_s}^2 \hat B_{B_s}  
\end{eqnarray}
At present, this new parameterization does not have a large effect on final 
results. It allows, however, to take better into account the uncertainty from 
the chiral extrapolation in lattice calculations of $f_{B_d}$. 

\subsection{Statistical methods}
The two statistical methods most often used in current UT analyses are the 
Bayesian~\cite{ref:noi1}-\cite{ref:noi3} and the
frequentistic~\cite{ref:loro1}-\cite{ref:loro2} approach. 

The Bayesian approach is based on the Bayes' theorem, which relates the 
probability of an event~$A$, given the event~$B$, to the probability of $B$ 
given~$A$:
\be
\label{eq:bayes1}
P(A|B) = P(B|A) \cdot P(A)\,/P(B) \,.
\ee
In the UT analysis, the event $A$ is represented by given values of the UT 
parameters, $\rhobar$ and $\etabar$, and of all the other parameters, denoted by
$\mathbf{x}$, ente\-ring the analysis (the Cabibbo angle, $|V_{cb}|$, the top 
quark mass, $\hat B_K$, etc.). The event $B$ represents instead the whole set of
experimental constraints (the measurements of $\epsilonk$, $\dmd$, $\snb$, 
etc.), denoted collectively by $\mathbf{c}$. The Bayes' theorem 
(\ref{eq:bayes1}) then allows one to compute the proba\-bi\-lity of given values
of $\rhobar$, $\etabar$ and $\mathbf{x}$ given the experimental constraints:
\be
\label{eq:bayes2}
f(\rhobar,\etabar,\mathbf{x}|\mathbf{c}) \propto f(\mathbf{c}|\rhobar,\etabar,
\mathbf{x}) \cdot f_0(\mathbf{x}) \, f_0(\rhobar,\etabar) \,.
\ee
The function $f_0(\mathbf{x})$ is the so called {\em a~priori} proba\-bility of 
$\mathbf{x}$. It expresses our knowledge on these parameters, coming for
instance from Lattice QCD calculations. As for the {\em a~priori} probability 
$f_0(\rhobar,\etabar)$, this is usually represented by a large flat 
distribution, since nothing is known about $\rhobar$ and $\etabar$ before the UT
analysis is performed. Integra\-ting both sides of Eq.~(\ref{eq:bayes2}) over 
all possible values of $\mathbf{x}$, finally leads to 
\be
\label{eq:bayes3}
f(\rhobar,\etabar|\mathbf{c}) \propto {\cal L}(\mathbf{c}|\rhobar,\etabar) 
\cdot f_0(\rhobar,\etabar) \,.
\ee
The function ${\cal L}(\mathbf{c}|\rhobar,\etabar)$ is the total likelihood.

\begin{figure}[t]
\vspace{-0.5cm}
\begin{center}
\includegraphics[width=3.5cm]{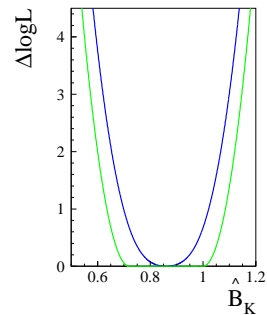}
\vspace{-1.0cm}
\caption{\it The $\Delta$-likelihood for $\hat B_K$ as obtained in the Bayesian 
(dark curve) and frequentistic (light curve) approach using $\hat B_K = 0.86 
\pm 0.06 \pm 0.14$.}
\label{fig:testbk}
\end{center}
\vspace{-1.0cm}
\end{figure}

There are several differences between the Bayesian and the frequentistic
approach, the most significant of which is the assumption within the latter that
an {\em a~priori} probability distribution function (p.d.f.) cannot be defined 
for theore\-ti\-cal parameters. As a consequence, in the frequentistic approach
the likelihood is defined in such a way that its ``theoretical" part does not 
contribute to the fit while the corresponding parameters take values within 
``allowed" ranges. In more recent versions of the frequentistic approach the 
above statement is applied in practice only to the systematic part of the 
uncertainties quoted for the theoretical parameters. Figure~\ref{fig:testbk} 
shows, as an example, the $\Delta$-likelihood for the parameter $\hat B_K$ as 
obtained in the Bayesian and in the frequentistic approach using $\hat B_K = 
0.86 \pm 0.06 \pm 0.14$.

As a member of a Bayesian collaboration, I do not need to specify which of the 
two approaches I do prefer. What I like about the Bayesian approach is that, 
being based on clear assumptions, it allows one to reach conclusions with a well
defined statistical meaning. This is not the case for the frequentistic 
approach, in which final results are usually quoted as having {\it at least} 
95\% of probability. The point I would most like to emphasize however is: 
irrespective of the preferred statistical method, the contribution to the 
likelihood coming from the knowledge of a given theoretical parameter, like 
$\hat B_K$ in Fig.~\ref{fig:testbk}, should be specified by the people who 
calculated it, rather than by those who perform the UT ana\-ly\-sis. In the 
specific example, it is our task in the lattice community to specify what we 
mean by ``$\hat B_K = 0.86 \pm 0.06 \pm 0.14$". In this way, the difference 
expressed by the two curves in Fig.~\ref{fig:testbk} would be resolved.

\subsection{Results for the UT parameters}
\begin{figure}[t]
\vspace{-0.5cm}
\begin{center}
\includegraphics[width=7.5cm]{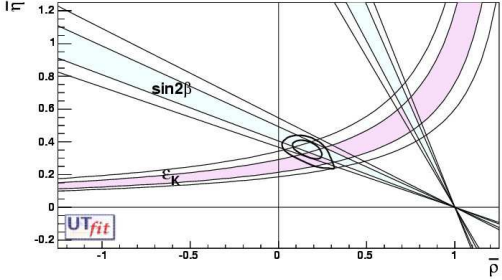}
\vspace{-1.2cm}
\caption{\it Comparison between the allowed regions for $\overline{\rho}$ and 
$\overline{\eta}$ (at 68\%, 95\% probability ranges) selected by measurements of
the CP conserving quantities 
and the bands derived from the measurements of CP violating quantities in the 
kaon ($\epsilonk$) and in the B ($\snb$) sectors.}
\label{fig:testcp}
\end{center}
\vspace{-0.9cm}
\end{figure}

%
The most significant test of the CKM mechanism of CP violation is the 
comparison between the ($\rhobar,\etabar$) region selected by the measurements 
of CP conserving quantities (semileptonic B decays and ${\rm B}^0-\overline{{
\rm B}^0}$ oscillations) and the regions selected by the measurements of CP 
violation in the kaon ($\epsilonk$) or in the B ($\snb$) sectors. This test is 
shown in Fig.~\ref{fig:testcp}.
It can be translated quantitatively by the comparison between the value of 
$\snb$ determined from the UT analysis when only CP conserving quantities 
are included (``sides" measurements) and the one obtained from the CP asymmetry 
measurement in $B \to J/\psi K_s$ decays:
\begin{eqnarray}
\snb = & 0.724 \pm 0.049 & ~~\rm {Sides~only} \nonumber \\
\snb = & 0.739 \pm 0.048 & ~~ B \to J/\psi K_s. 
\label{eq:sin2beta}
\end{eqnarray}
The excellent agreement between these values illustrates the consistency of the 
SM in describing CP violation phenomena in terms of one single parameter 
$\etabar$. 

\begin{figure}[t]
\vspace{-0.5cm}
\begin{center}
{\includegraphics[height=3.5cm]{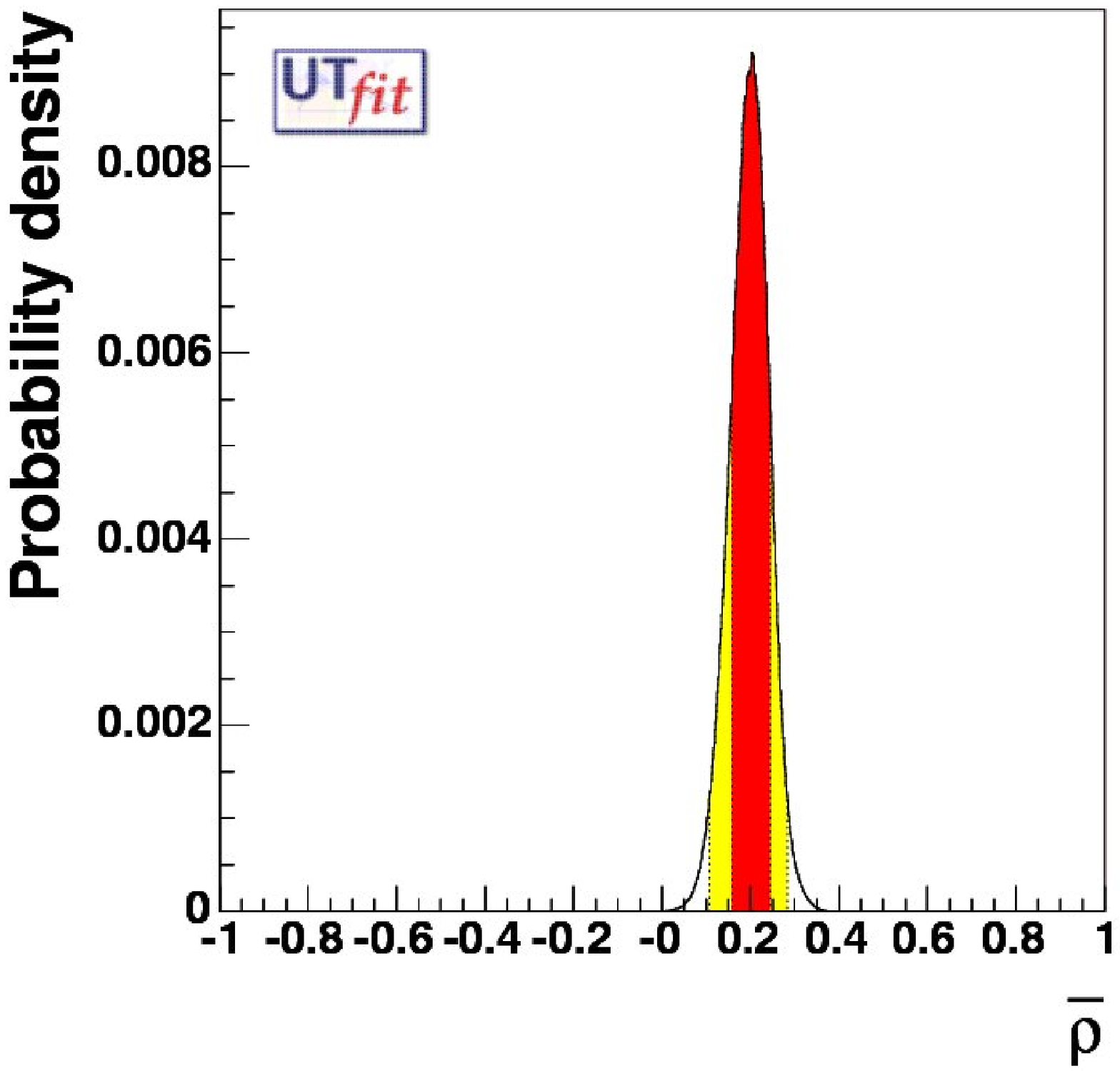}}
{\includegraphics[height=3.5cm]{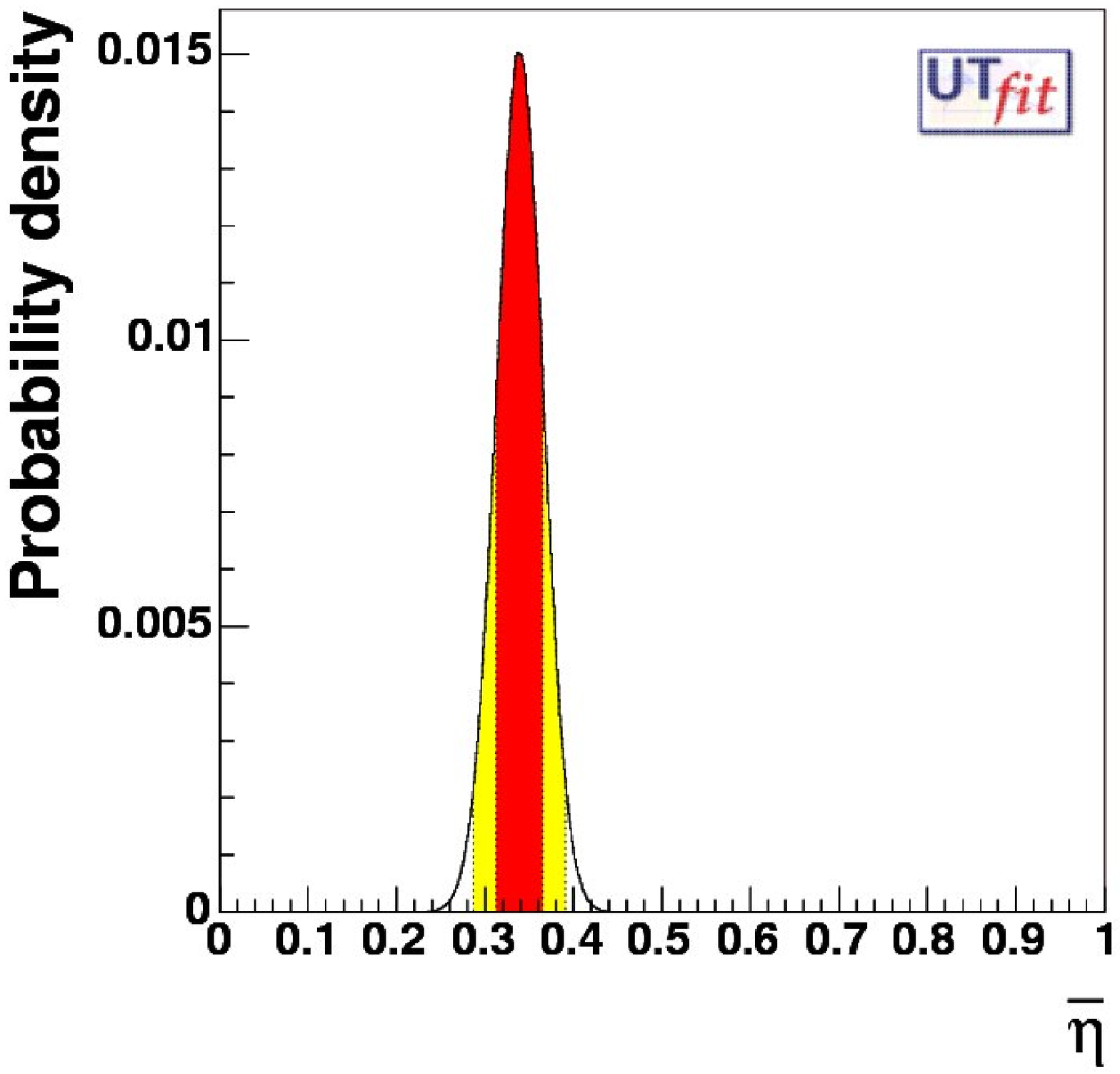}}
{\includegraphics[height=3.5cm]{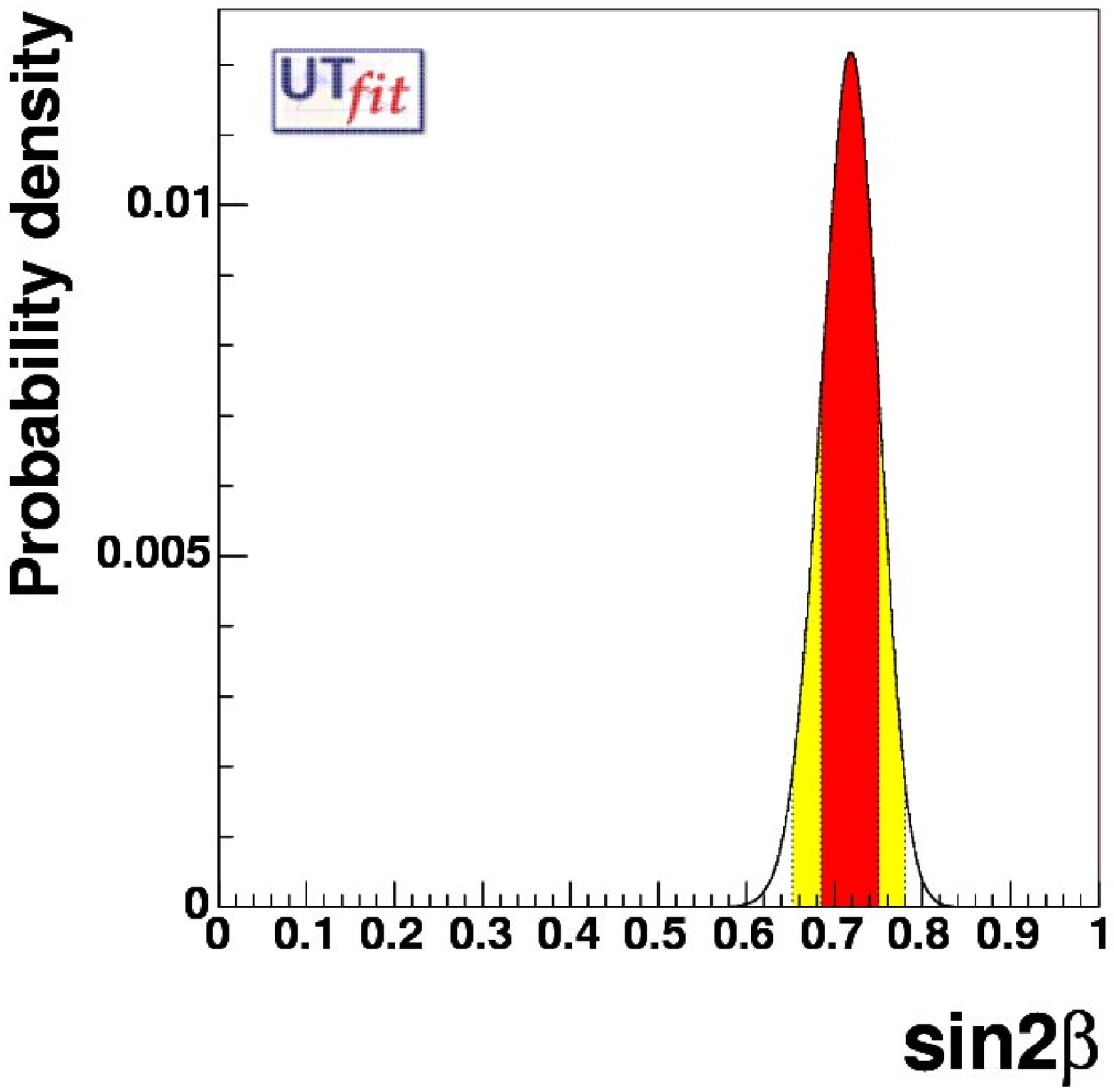}}
{\includegraphics[height=3.5cm]{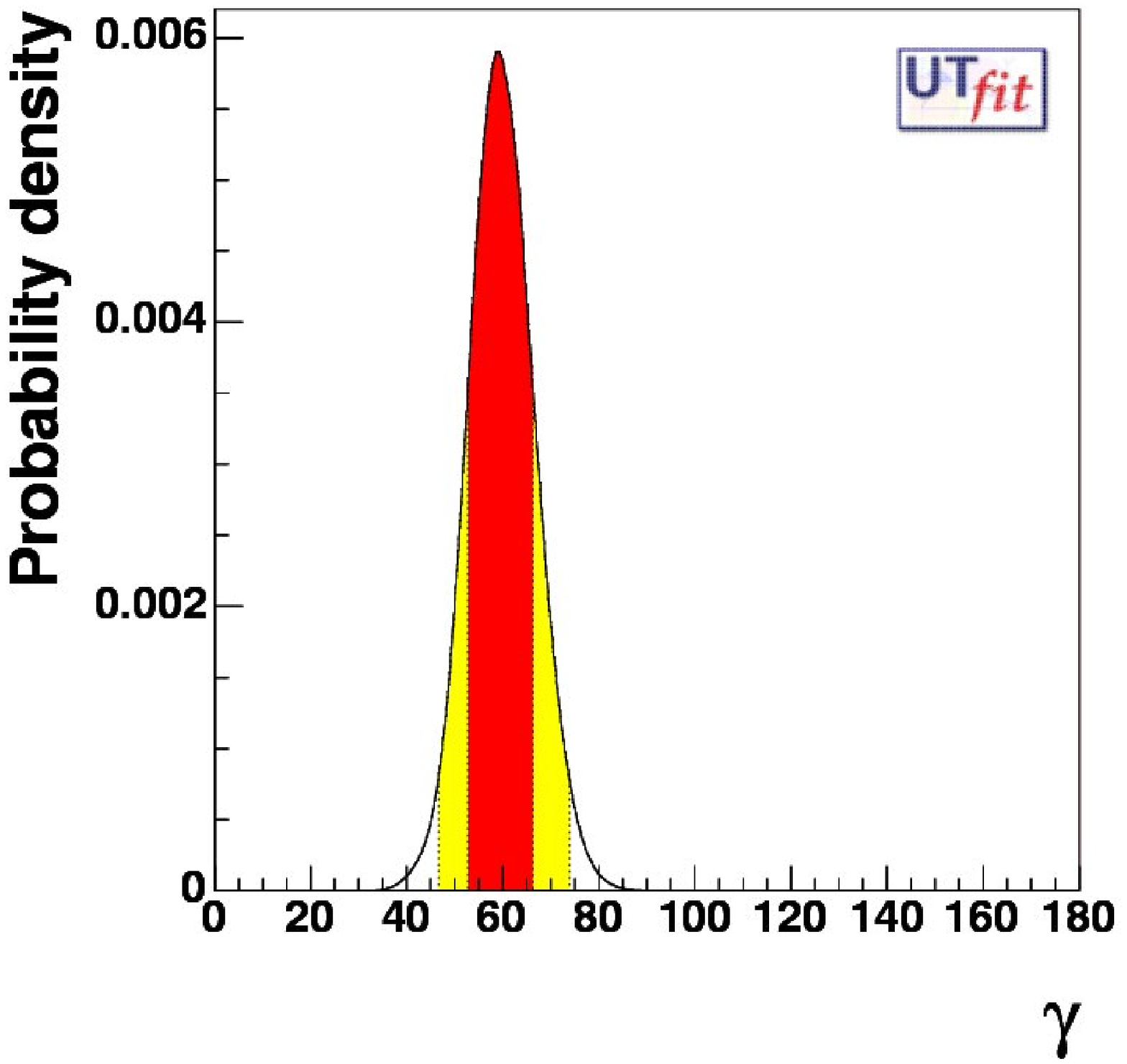}}
\vspace{-1.3cm}
\caption{\it P.d.f.'s for $\rhobar$, $\etabar$, $\snb$ and $\gamma$. The red 
(dark) and the yellow (light) zones correspond respectively to 68\% and 95\% of 
probability.}
\label{fig:1dim}
\end{center}
\vspace{-1.0cm}
\end{figure}

\begin{figure}[thb]
\vspace{-0.3cm}
\begin{center}
\includegraphics[width=7.5cm]{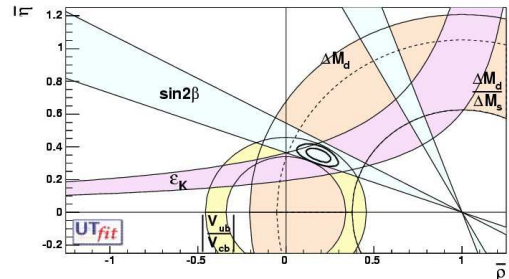}
\vspace{-1.5cm}
\caption{ \it Allowed regions (contours at 68\% and 95\% probability) for 
$\rhobar$ and $\etabar$. The bands show the 95\% probability regions selected 
from the various constraints.}
\label{fig:rhoeta}
\end{center}
\vspace{-0.8cm}
\end{figure}

The results for the main UT parameters obtained from the standard analysis are 
presented in Table~\ref{tab:1dim}. Figures~\ref{fig:1dim} and \ref{fig:rhoeta} 
show respectively some of the corresponding p.d.f.'s and the selected region in 
the ($\rhobar,\etabar$) plane.
\begin{table}[t]
\caption {\it Values of the main UT parameters obtained from the standard
analysis.}
\label{tab:1dim} 
\begin{center}
\begin{tabular}{cc}
\hline\hline
    Parameter          &      \\ \hline 
$\overline {\eta}$     & 0.348  $\pm$ 0.028   \\
$\overline {\rho}$     & 0.172  $\pm$ 0.047   \\
      $\snb$           & 0.725  $\pm$ 0.033   \\
      $\sna$           & -0.16  $\pm$ 0.26    \\
$\gamma[^{\circ}$]     & 61.5   $\pm$ 7.0     \\
\hline\hline
\end{tabular} 
\end{center}
\vspace{-1.0cm}
\end{table}

Another interesting prediction of the UT ana\-ly\-sis concerns the mass 
difference $\dms$. This prediction is obtained by removing from the fit the 
experimental information coming from ${\rm B}^0_s-\overline{\rm {B}^0_s}$ 
oscillations (present analyses at LEP/SLD have established a sensitivity of 
18.3~ps$^{-1}$ and they show a higher probability region for a positive signal 
around 17.5 ps$^{-1}$). The prediction is
\be
\label{eq:dmsnodms}  
\dms = (21.1 \pm 3.1) ~ {\mathrm ps}^{-1} \,.
\ee
It is also interesting to quote the value obtained when the experimental 
information on ${\rm B}^0_s-\overline{\rm {B}^0_s}$ oscillations is included in 
the analysis,
\be  
\dms = (18.3 \pm 1.6) ~ {\mathrm ps}^{-1} ~~~ (\mathrm{including}~\dms)\,.
\ee
Once an accurate measurement of $\dms$, expected from the TeVatron in the near 
future, will become available, the prediction of Eq.~(\ref{eq:dmsnodms}) will 
provide a crucial test of the SM.

\subsection{Hadronic parameters}
A study of particular interest from the Lattice QCD point of view is the 
determination of the hadronic parameters from the UT analysis. This study 
consists of removing from the analysis, input information coming from Lattice 
QCD, namely the estimates of $\fbssqbs$, $\xi$ and $\hat{B}_K$, by treating 
these quantities as free parameters in the fit. In this way, the result obtained
from the UT fit for a given hadronic parameter can be compared to the 
corresponding lattice determination. Table~\ref{tab:nonptsumm1} shows the 
results of the fit when one hadronic parameter at a time is treated as a free 
parameter. The current Lattice QCD determinations are shown for comparison.  
\begin{table}[t]
\caption {\it Values of the non-perturbative QCD parameters as obtained from 
the UT analysis. The current Lattice QCD determinations are also shown for 
comparison.}
\label{tab:nonptsumm1} 
\begin{center}
\begin{tabular}{c|c|c}
\hline\hline
Parameter &  From UT fit  &  From Lattice QCD \\
\hline
$\hat{B}_K$   & 0.65 $\pm$ 0.10 & 0.86 $\pm$ 0.06 $\pm$ 0.14      \\ \hline
$\fbssqbs$    & 263  $\pm$ 14 MeV  & 276  $\pm$ 38 MeV         \\ \hline
$\xi$         & 1.13$^{+0.12}_{-0.09}$ & 1.24 $\pm$ 0.04 $\pm$ 0.06 \\ \hline
\hline\end{tabular} 
\end{center}
\vspace{-1.0cm}
\end{table}

Some conclusions can be drawn. The precision on $\fbssqbs$ obtained from the 
UT fit has an accuracy which is better than the current Lattice QCD estimate. 
This proves that the standard CKM fit is, in practice, weakly dependent on the 
theoretical uncertainty on this parameter. The UT fit result for $\hat{B}
_K$ indicates that values of $\hat{B}_K$ smaller than $0.45$ are excluded at 
99\% probability. The accuracy achieved in the estimate of $\hat{B}_K$ from the 
fit is at the level of 15\%, comparable to the one obtained from Lattice QCD. 
On the other hand, the best determination of the parameter $\xi$ comes at 
present from the lattice.

\subsection{Impact of improved determinations}
%
Before concluding this section I would like to illustrate to what extent the UT 
analysis could benefit from more accurate lattice calculations in the next few
years. I assume that the uncertainties in the values of the lattice input 
parameters will be reduced as indicated in Table~\ref{table:future}, to be 
compared with current uncertainties presented in Table~\ref{tab:inputs}.
\begin{table}[t]
\caption {\it Values of the input parameters used in the UT analysis in the 
projection for the next years.}
\label{table:future} 
\begin{center}
\begin{tabular}{c|c}
\hline\hline
Parameter    & Next years \\ \hline 
$\hat{B}_K$  & $0.86 \pm 0.06$ \\
$\fbssqbs$   & $276 \pm 14~\MeV$  \\
$\xi$        & $1.24 \pm 0.04$  \\
$\left|V_{ub}\right|$(excl.) & $(33.0 \pm 2.4) \times 10^{-4}$  \\
$\snb$       & $0.739 \pm  0.021$  \\
\hline\hline
\end{tabular} 
\end{center}
\vspace{-1.0cm}
\end{table}
A discussion on the extent to which such a projection is realistic is beyond the
scope of this paper. Besides reducing the uncertainties in the hadronic 
parameters, I take into account the impact of a more accurate experimental 
determination of $\snb$ from $B \to J/\psi K_s$ decays.

\begin{figure}[t]
\vspace{-0.3cm}
\begin{center}
\includegraphics[width=7.5cm]{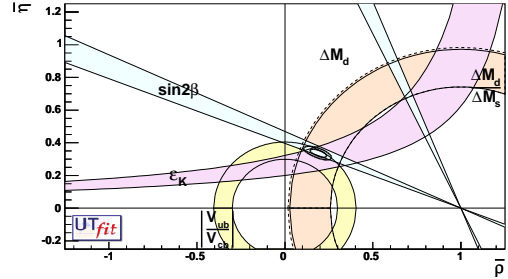}
\vspace{-1.5cm}
\caption{ \it The contours at 68\% and 95\% probability in the 
($\rhobar$,$\etabar$) plane as obtained in the projection for the next years.}
\label{fig:rhoetafuture}
\end{center}
\vspace{-1.0cm}
\end{figure}

The region in the ($\rhobar$,$\etabar$) plane selected by the analysis with the
projection for the near future is shown in Fig.~\ref{fig:rhoetafuture}, which
is to be compared with the present determination shown in Fig.~\ref{fig:rhoeta}.
Quantitatively, the effect is a reduction of the uncertainties on both $\rhobar$
and $\etabar$ by approximately 40\% (from 27\% to 17\% for $\rhobar$ and from 
8\% to 5\% for $\etabar$). In achieving this result, the assumption of a 5\% 
accuracy reached in the determination of $\fbssqbs$ plays a minor role since, 
as discussed before, the value of this parameter is already quite constrained by
the UT fit. On the other hand, the precision in the determination of $\fbssqbs$ 
has a crucial impact on the prediction of the mass difference $\dms$. The value 
obtained from the projection for the future is
\be
\label{eq:dmsfuture}  
\dms = (19.7 \pm 1.8) ~ {\mathrm ps}^{-1} \,,
\ee
to be compared with the present determination quoted in Eq.~(\ref{eq:dmsnodms}).

\section{CONCLUSIONS}
Flavor physics in the quark sector has entered its mature age. The precision
reached in the determination of the fundamental parameters, namely quark masses
and CKM matrix elements, has significantly improved. The mixing between the 
first two generations of quarks is accurately determined, and recent 
experimental results on $K_{\ell3}$ decays, combined with the precise 
theoretical determination of the relevant form factor, provide a value of 
$|V_{us}|$ in excellent agreement with the unitarity prediction. 
The UT parameters, which define the mixing between the first and the third 
generations, are determined with good precision. A crucial test of the CKM 
mechanism of CP vio\-la\-tion has already been performed, namely the comparison 
between the UT parameters as determined from CP conserving quantities 
(semileptonic B decays and $\Bo-\Bob$ oscillations), and the measurements of CP 
violation in the kaon ($\epsilon_K$) and in the B (sin2$\beta$) sectors. The 
agreement is excellent. Such a test could not have been performed without the 
essential contribution of Lattice QCD calculations.

Though I did not have time to discuss them in this talk, it should be mentioned
that measurements of non-leptonic B meson decays at the B-Factories have already
started to have an impact on the UT analysis~\cite{ref:noi3}. In the next years,
these measurements will allow tests of the SM in the flavor sector to an 
accuracy up to the per cent level. A crucial role will be also played by 
precision charm physics. Accurate measurements of leptonic and semileptonic 
D-meson decay rates will be available soon from CLEO-c~\cite{shipsey}, allowing 
for precise determinations of the CKM parameters $V_{cs}$ and $V_{cd}$. In this
important research, Lattice QCD calculations are expected to provide the 
necessary theoretical support.

\section*{ACKNOWLEDGMENTS}
I wish to thank C.~Bernard, P.~Gambino, L.~Giusti, G.~Isidori, M.~Wingate,
S.~Sharpe and I.~Shipsey for useful discussions. I am grateful to A.~Kronfeld 
and E.~Freeland for a careful reading and for valuable suggestions on the 
manuscript. I am also grateful to all my friends and colleagues of the 
SPQ$_\mathrm{CD}$R and UTFit Collaborations for an enjoyable, fruitful and 
longstanding collaboration.

\end{document}